\documentclass[iop,numberedappendix]{emulateapj}
\usepackage{natbib}
\usepackage{amsmath}
\newcommand{\gps}{\ensuremath{g_{\rm P1}}}
\newcommand{\rps}{\ensuremath{r_{\rm P1}}}
\newcommand{\ips}{\ensuremath{i_{\rm P1}}}
\newcommand{\ipsl}{\ensuremath{i_{\rm P1,lim}}}
\newcommand{\zps}{\ensuremath{z_{\rm P1}}}
\newcommand{\zpsa}{\ensuremath{z_{\rm P1-aper}}}
\newcommand{\yps}{\ensuremath{y_{\rm P1}}}
\newcommand{\ypsa}{\ensuremath{y_{\rm P1-aper}}}

\newcommand{\PS}{\protect \hbox {Pan-STARRS1}}

\newcommand{\isdss}{\ensuremath{i_{\rm SDSS}}}

\newcommand{\Jukidss}{\ensuremath{J_{\rm UKIDSS}}}

\newcommand{\Jtmass}{\ensuremath{J_{\rm 2M}}}
\newcommand{\Htmass}{\ensuremath{H_{\rm 2M}}}
\newcommand{\Ktmass}{\ensuremath{K_{\rm 2M}}}

\newcommand{\igrond}{\ensuremath{i_{\rm GROND}}}
\newcommand{\zgrond}{\ensuremath{z_{\rm GROND}}}
\newcommand{\Jgrond}{\ensuremath{J_{\rm GROND}}}
\newcommand{\Hgrond}{\ensuremath{H_{\rm GROND}}}

\newcommand{\intt}{\ensuremath{I_{\rm EFOSC2}}}
\newcommand{\zntt}{\ensuremath{Z_{\rm EFOSC2}}}

\newcommand{\zotk}{\ensuremath{z_{\rm O2000}}}
\newcommand{\Yotk}{\ensuremath{Y_{\rm O2000}}}
\newcommand{\Jotk}{\ensuremath{J_{\rm O2000}}}

\newcommand{\lya}{Ly\ensuremath{\alpha}}
\newcommand{\nv}{N\ensuremath{\,\textsc{v}}}
\newcommand{\oi}{O\ensuremath{\,\textsc{i}}\,+\,Si\ensuremath{\,\textsc{ii}}} 

\newcommand{\SIii}{Si\ensuremath{\,\textsc{ii}}}
\newcommand{\mgii}{Mg\ensuremath{\,\textsc{ii}}}
\newcommand{\mgi}{Mg\ensuremath{\,\textsc{i}}}

\newcommand{\feii}{Fe\ensuremath{\,\textsc{ii}}}
\newcommand{\civ}{C\ensuremath{\,\textsc{iv}}}
\newcommand{\cii}{C\ensuremath{\,\textsc{ii}}}

\newcommand{\siivpoiv}{Si\ensuremath{\,\textsc{iv}\,+\,}O\ensuremath{\,\textsc{iv]}}}

\def\msun{{\rm\,M_\odot}}

\def\deg{^\circ}

\def\sec{\hbox{\arcsec\hskip-3pt .}}

\def\s{\ifmmode \widetilde \else \~\fi}
\def\={\overline}

\def\spose#1{\hbox to 0pt{#1\hss}}

\def\lta{\mathrel{\spose{\lower 3pt\hbox{$\mathchar"218$}}
     \raise 2.0pt\hbox{$\mathchar"13C$}}}
\def\gta{\mathrel{\spose{\lower 3pt\hbox{$\mathchar"218$}}
     \raise 2.0pt\hbox{$\mathchar"13E$}}}
\def\simlt{\lower.5ex\hbox{\ltsima}}
\def\gtsima{$\; \buildrel > \over \sim \;$}
\def\simgt{\lower.5ex\hbox{\gtsima}}
\def\s{\ifmmode \widetilde \else \~\fi}
\def\={\overline}

\def\lta{\mathrel{\spose{\lower 3pt\hbox{$\mathchar"218$}}
     \raise 2.0pt\hbox{$\mathchar"13C$}}}
\def\gta{\mathrel{\spose{\lower 3pt\hbox{$\mathchar"218$}}
     \raise 2.0pt\hbox{$\mathchar"13E$}}}
\def\Dt{\spose{\raise 1.5ex\hbox{\hskip3pt$\mathchar"201$}}}    
\def\dt{\spose{\raise 1.0ex\hbox{\hskip2pt$\mathchar"201$}}}    

\def\dotsfill{\leaders\hbox to 1em{\hss.\hss}\hfill}

\shorttitle{Discovery of $z\sim 6$ quasars from \PS}
\shortauthors{Ba\~{n}ados et al.}
\begin{document}
\newcommand{\Qa}{\protect \hbox {PSO J340.2041--18.6621}}
\newcommand{\Qb}{\protect \hbox {PSO J007.0273+04.9571}}
\newcommand{\Qc}{\protect \hbox {PSO J037.9706--28.8389}}
\newcommand{\Qd}{\protect \hbox {PSO J187.3050+04.3243}}
\newcommand{\Qe}{\protect \hbox {PSO J213.3629--22.5617}}
\newcommand{\Qf}{\protect \hbox {PSO J183.2991--12.7676}}
\newcommand{\Qg}{\protect \hbox {PSO J210.8722--12.0094}}
\newcommand{\Qh}{\protect \hbox {PSO J045.1840--22.5408}}

\newcommand{\Qeric}{\protect \hbox {PSO J215.1514--16.0417}}

\title{Discovery of eight $z\sim 6$ quasars from Pan-STARRS1}

\author{
E.~Ba\~{n}ados \altaffilmark{1},
B.P.~Venemans \altaffilmark{1},
E.~Morganson \altaffilmark{1},
R.~Decarli \altaffilmark{1},
F.~Walter \altaffilmark{1},
K.C.~Chambers \altaffilmark{2},
H-W.~Rix \altaffilmark{1},
E.P.~Farina\altaffilmark{1},
X.~Fan\altaffilmark{3},
L.~Jiang\altaffilmark{4},
I.~McGreer \altaffilmark{3},
G.~De~Rosa\altaffilmark{5},
R.~Simcoe\altaffilmark{6},
A.~Wei\ss\altaffilmark{7},
P.~A. Price\altaffilmark{8},  
J.~S.~Morgan\altaffilmark{2}, 
W. S. Burgett\altaffilmark{2}, 
J. Greiner\altaffilmark{9	},
N. Kaiser\altaffilmark{2}, 
R.-P. Kudritzki\altaffilmark{2}, 
E. A. Magnier\altaffilmark{2}, 
N. Metcalfe\altaffilmark{10}, 
C. W. Stubbs\altaffilmark{11}, 
W.~Sweeney \altaffilmark{2}, 
J. L. Tonry\altaffilmark{2},
R. J. Wainscoat\altaffilmark{2}, 
 and
C. Waters\altaffilmark{2} 
}

\altaffiltext{1}{Max Planck Institut f\"ur Astronomie, K\"onigstuhl 17, D-69117, Heidelberg, Germany}%
\email{banados@mpia.de}
\altaffiltext{2}{Institute for Astronomy, University of Hawaii, 2680 Woodlawn Drive, Honolulu, HI 96822, USA}
\altaffiltext{3}{Steward Observatory, The University of Arizona, 933 North Cherry Avenue, Tucson, AZ 85721--0065, USA}
\altaffiltext{4}{School of Earth and Space Exploration, Arizona State University, Tempe, AZ 85287, USA}
\altaffiltext{5}{Department of Astronomy, The Ohio State University, 140 West 18th Avenue, Columbus, OH 43210, USA}
\altaffiltext{6}{MIT-Kavli Center for Astrophysics and Space Research, 77 Massachusetts Avenue, Cambridge, MA, 02139, USA}
\altaffiltext{7}{Max-Planck-Institut f\"{u}r Radioastronomie, Auf dem H\"{u}gel 69 D-53121 Bonn, Germany}
\altaffiltext{8}{Department of Astrophysical Sciences, Princeton University, Princeton, NJ 08544, USA}
\altaffiltext{9}{Max-Planck-Institut f\"ur extraterrestrische Physik, Giessenbachstrasse 1, 85748 Garching, Germany}
\altaffiltext{10}{Department of Physics, Durham University, South Road, Durham DH1 3LE, UK}
\altaffiltext{11}{Department of Physics, Harvard University, Cambridge, MA 02138, USA}

\begin{abstract}
 High-redshift quasars are currently the only probes of the growth of supermassive 
 black holes and potential tracers of structure evolution at early cosmic time.  
 Here we present our candidate selection criteria from the Panoramic  Survey 
 Telescope \& Rapid Response System 1 and follow-up strategy to discover
 quasars in  the redshift range $5.7 \lesssim  z \lesssim  6.2$.
With this strategy we discovered eight new $5.7 \leq z \leq 6.0$ quasars, increasing the number 
of known quasars at $z>5.7$ by more than 10 \%.  
 We additionally recovered 18 previously known quasars. 
The eight quasars presented here span a large range of luminosities 
($-27.3 \leq M_{1450} \leq -25.4$; $19.6 \leq  \zps \leq 21.2$) and are remarkably heterogeneous 
in their spectral features: half of them show bright emission lines whereas the other half show a weak or no Ly$\alpha$ emission line 
(25\% with rest-frame equivalent width of the Ly$\alpha +$N\textsc{v} line lower than 15\AA).
We find a larger fraction of weak-line emission quasars than in lower redshift studies.
This may imply that the weak-line quasar 
population at the highest redshifts could be more abundant than previously thought.
However, larger samples of quasars are needed to increase the statistical significance of this finding.
\end{abstract}


\vfil
\eject
\clearpage

\section{INTRODUCTION}
\label{sec:intro}

High-redshift quasars provide us with unique information about the      
evolution of supermassive black holes (SMBHs), their host galaxies, and the intergalactic
medium (IGM) at early cosmic time. Over the last decade, numerous 
studies have established a sample of roughly $60$ quasars at $5.5 < z < 7.1$,
mostly discovered using optical surveys such as the Sloan Digital Sky Survey \citep[SDSS;][]{fan01, fan03,fan04, fan06a}  and
the Canada-France High-$z$ Quasar Survey \citep[CFHQS;][]{will05a, will07, will09, will10, will10b}.
Among the key results from these studies are the existence of
SMBHs less than a gigayear after the big bang \citep[e.g.,][]{will03, mor11, der13}, and the presence of near
complete Gunn--Peterson absorption, indicating a rapid increase in the
IGM neutral fraction and the end of reionization at $z \sim 6$ \citep{fan06c}.
These findings strongly suggest that fundamental changes are
happening in the IGM at $6 \lesssim z \lesssim 7$. In addition, the initial formation
and growth of these SMBHs is, however, not fully understood. In order to further study 
this important era in the history of the universe and better 
constrain the formation and evolution of early structures and SMBH, the discovery
and characterization of a significant sample of quasars in this 
redshift range is crucial.

The Panoramic Survey Telescope \& Rapid Response System 1 \citep[\PS, PS1;][]{kai02, kai10} $3\pi$
is surveying all the sky above declination $-30\deg$ in the filters \gps, 
\rps, \ips, \zps, and \yps\ \citep{stu10, ton12}.
The PS1 $3\pi$ survey maps yearly each region of the sky in two sets of pairs per filter. 
Each pair is taken during the same night. The second pair 
for \gps, \rps, and \ips\ are obtained during the same lunation, whereas  for \zps\ and \yps\ are observed approximately 5--6 months 
later.  There are some regions
of the sky that are not covered or have less coverage (i.e., 0 or 2 exposures per year) 
than the average survey mostly due to weather restrictions (especially for \gps, \rps, and \ips\
bands because their observation pairs are taken within the same month) or areas falling 
into the camera chip gaps.

PS1 presents an excellent opportunity to perform high-redshift quasar 
searches for three reasons: (1) it covers a larger area than previous high-redshift quasar surveys, especially in the southern hemisphere; (2) it goes significantly deeper than SDSS
in the reddest bands where $z \sim 6$ quasars are visible (current $5\sigma$ median limiting magnitudes are
$\gps=22.9$, $\rps=22.8$, $\ips=22.6$, $\zps=21.9$, $\yps=20.9$); and 
(3) the additional $y$-band \citep[$\lambda_{\rm eff} = 9620$ \AA; FWHM$=890$ \AA;][]{ton12} enables the search for luminous quasars beyond the SDSS limit,
$z>6.5$. In early 2013, PS1 produced its first internal release of the $3\pi$ stacked catalog (PV1), 
which is based on the co-added PS1 exposures \citep[see][]{met13}.
This is the catalog used for the present work and it includes data obtained primarily during the period 2010 May--2013 March.

In \cite{morg12}, we presented our preliminary selection method for 
our quasar search when the stacked catalog was not available and
reported the discovery of the first PS1 high-redshift quasar (\Qeric\ \footnote{We have slightly modified the name published in \cite{morg12}
from PSO J215.1512--16.0417 to \mbox{PSO J215.1514--16.0417} based on the stacked catalog coordinates.} at $z=5.73$).
In this paper, we refined our
selection criteria and made use of the deeper and more complete information in the new PS1 stacked catalog. 
Our aim  was to discover quasars at $ z\sim 6$ while minimizing the contamination of foreground objects.
This was possible by focusing the search on the redshift range $5.7 \lesssim z \lesssim 6.2$ where quasars and stars
are best differentiated in PS1 colors (for details see Section \ref{sec:selection}). In this work we present our strategy and the first discoveries
of this search.

The paper is organized as follows. In Section 2, we present our 
color selection procedures for $5.7 \lesssim z \lesssim 6.2$ quasars, including the initial selection from
the PS1 stacked catalog and the follow-up optical and near-infrared (NIR) photometry. In Section 3, we present the discovery of eight 
new $z\sim 6$ quasars and discuss their spectral properties. In Section \ref{sec:known_qsos}, we estimate our completeness by studying whether other quasars 
from the literature were or were not part of our list of candidates. 
We discuss  how typical the weak-line emission quasars discovered in this work are in Section \ref{sec:wlq}
and summarize our results in Section \ref{sec:summary}. Magnitudes throughout the paper are given in the AB system, except
when referring to Two Micron All Sky Survey (2MASS) and Wide-field Infrared Survey Explorer (\textit{WISE}) magnitudes which are in the Vega system, unless otherwise stated.
We employ a cosmology with $H_0 = 69.3 \,\mbox{km s}^{-1}$ Mpc$^{-1}$, $\Omega_M = 0.29$, and $\Omega_\Lambda = 0.71$ \citep[][]{wmap9}.

\section{Selection of Quasar Candidates}
\label{sec:selection}
Quasars at redshift $z\gtrsim 5.7$ 
are observationally characterized by their very red $i-z$ color and blue continuum (i.e., blue $z-y$ color).
They are very faint or completely undetected in 
the $i$ band due to the optically thick Ly$\alpha$ forest at these redshifts,
causing most of the light coming from wavelengths $\lambda_{{\rm rest}} < 1216$ \AA\ 
to be absorbed.

To determine the optimum selection criteria, we have taken the composite quasar spectrum from \cite{dec10}, 
which consists of 96 bright quasars at $0<z<3$. We chose this template because at red wavelengths it is less affected by 
host galaxy contamination than other composite spectra in the literature \citep[e.g.,][]{fra91,van01} and it samples 
the rest frame wavelengths of $5.7 < z <7.2$ quasars in the PS1 filters. For $\lambda_{{\rm rest}} < 1010 $ \AA\ we replaced the 
template with a flat $f_\lambda$, normalized to the UV continuum blueward of \lya. Then, we applied the intergalactic attenuation correction
from \cite{mei06}, tuning the optical depth, $\tau_{\rm IGM}$, to match the transmitted flux blueward of \lya\ in the $z \sim 6$ SDSS
composite spectrum from \cite{fan06a}. 
Figure \ref{fig:selection} shows the expected track of our quasar template through the $\ips - \zps$
and $\zps - \yps$  colors as the redshift is increased from $z=5.0$ to $z=6.5$.
We are also interested in the location of our potential contaminants in this diagram.
Since the observational color scatter for brown dwarfs is significant, we decided to compare with real PS1 colors of known brown dwarfs. 
We cross-matched the ultracool dwarfs presented in \cite{dup12} with the PS1 stacked
catalog, taking the closest match within a $ 5\arcsec$ radius with proper motion measurements taken into account. There were 126 matches that had
measurements in the \ips, \zps, and \yps\ bands, and with a signal-to-noise ratio (S/N) $> 10$ in the \zps\ band and S/N $> 5$ in the \yps\ band. Their PS1 colors
 are represented by stars in Figure \ref{fig:selection}.

In this work we focused on the upper-left region of the color--color diagram in the right
panel of Figure \ref{fig:selection} for two reasons: (1) the bulk of the 
$z\sim 6$ quasar population is located there and (2) this is the region with the least brown dwarf contamination in this color--color diagnostic.

Because high-redshift quasars are very rare and because of the high number of other sources (or artifacts) that mimic
high-redshift quasar colors, we cleaned our sample in several steps:

\begin{enumerate}
 \item We selected initial high-redshift quasar candidates from the PV1 PS1 stacked database (see Section \ref{ps1_selection}).
 \item At the position of each candidate, we applied forced aperture photometry in the stacked images in order to corroborate  
   the catalog colors (see Section \ref{sec:stack_phot}).
 \item This was followed by forced photometry in all single-epoch $z$-band images to remove artifacts (see Section \ref{sec:single_phot}).
 \item We then matched the candidate list to the 2MASS \citep[][]{skr06}, SDSS DR8 \citep{aih11}, and the
 UKIRT Infrared Deep Sky Survey \citep[UKIDSS;][]{law07} to eliminate contaminants that are evident when using the extra information
 provided by these surveys (see Section \ref{sec:cross_surveys}).
 \item We then cross-matched the remaining candidates with known quasars (see Section \ref{sec:cross-match}) and visually
inspected the stacked and single-epoch stamps to ensure that they are real.
  \item We then obtained optical and NIR follow-up photometry (see Section \ref{sec:img_follow}).
 \item Finally, we obtained spectra to confirm the nature and redshifts of the remaining candidates (see Section \ref{sec:spc_follow}). 
 \end{enumerate}

\subsection{Selection from Pan-STARRS1 Catalog } \label{ps1_selection}
 
We excluded the Galactic plane ($\left| b \right| < 20\deg$), and
M31 ($7\deg< \,$R.A.$ <14\deg$; $37\deg< \,$Decl.$ <43\deg$) from our search.
The catalog used in this work had no data between $92\deg <\,$R.A.$ <132\deg$. 
After removing these regions, the effective area of our survey is $\sim 20,000$ square degrees
or $\sim 1.9\,\pi$ steradian.

For our candidates, we required coverage in at least
the three redder bands (\ips, \zps, and \yps) and that more than $85\%$ of the expected
point-spread function (PSF)-weighted flux in these bands was located in valid pixels (i.e., that the PS1 catalog entry has 
{\tt PSF\_QF} $> 0.85$). In order to estimate the fraction of objects missed by requiring coverage in these three bands and/or by being located in bad pixels,
we cross-matched the PS1 catalog to sources in 2MASS with S/N $>10$ and fainter than 14th magnitude in all the 2MASS bands.
The magnitude cut is intended to avoid objects that would be saturated in the PS1 catalog. After applying the same coverage and pixel criteria we estimate
that the percentage of missed objects is $\sim 7\%$.
We also excluded those measurements for which the Image Processing Pipeline \citep[][]{mag06, mag07} flagged the result as suspicious
(see Table \ref{table:flags} in Appendix \ref{ap:ps1_flags} for details).

\begin{figure*}
\epsscale{1.0}
\plotone{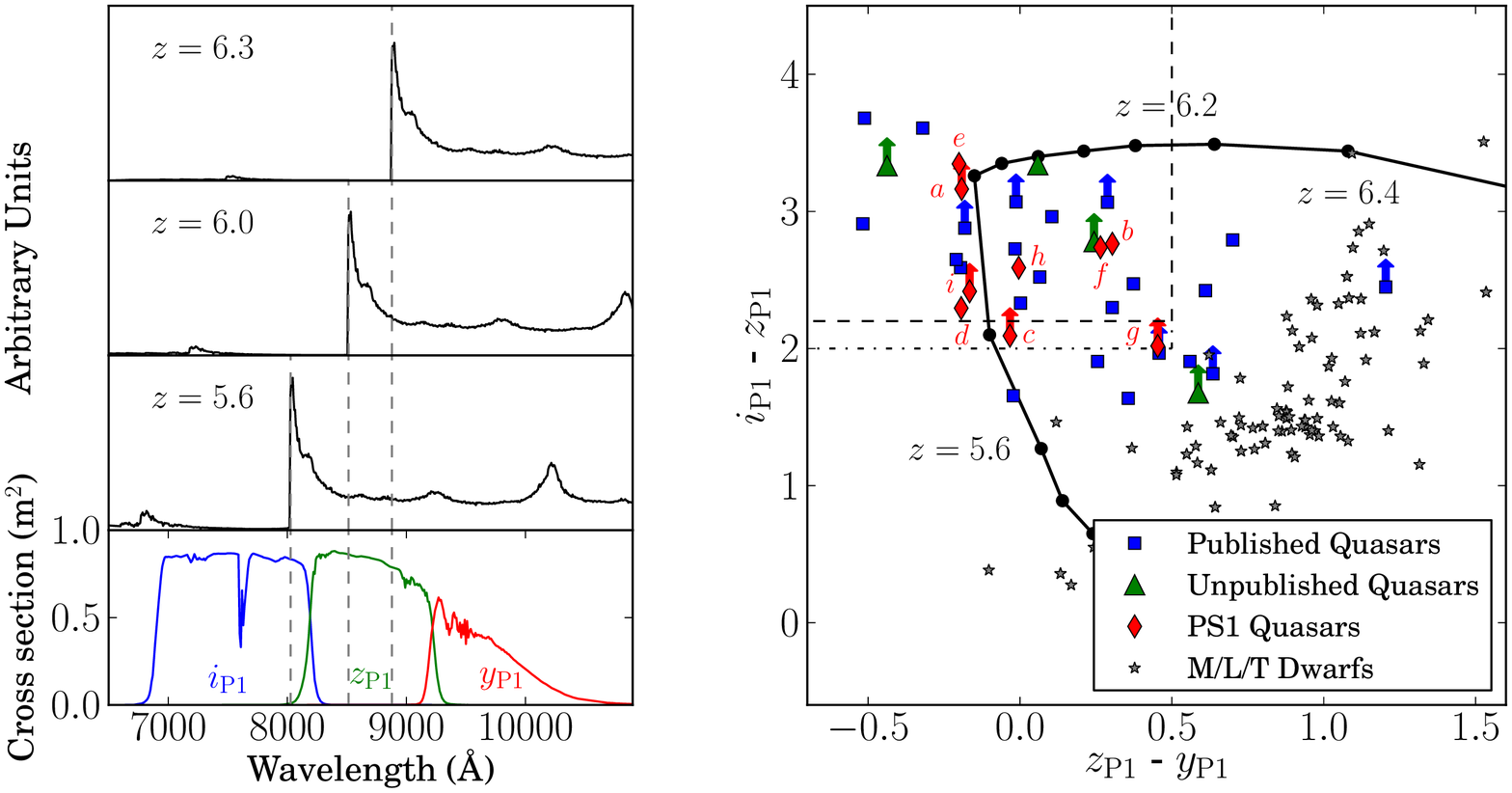}
\caption{
Left: the bottom panel shows 
the PS1 capture cross section in units of m$^2$ e$^{-1}$ photon$^{-1}$ for the 
\ips, \zps, and \yps\ bands \citep{ton12}.
The three upper panels show the composite quasar spectrum
from \cite{dec10} (with the intergalactic medium correction 
from Meiksin 2006) redshifted to $z=5.6$, $z=6.0$, and $z=6.3$,
from bottom to top. The gray dashed lines show the wavelength of
the \lya\ line at each redshift.
Right: color--color diagram showing the criteria used 
to select quasar candidates (long-dashed line, upper left corner).
The dot-dashed line is used for candidates with upper limits 
in the \ips\ band. The thick black line shows the expected color of
the quasar template from \cite{dec10} redshifted from $z=5.0$ to $z=6.5$ in steps of $\Delta z = 0.1$ (see left panel). The M/L/T dwarfs from \cite{dup12} that have a PS1 counterpart
are shown with stars. 
Blue squares are published quasars at $5.7 <z< 6.4$ satisfying our 
S/N and coverage criteria. Fourteen of them were included in 
our candidate list. Green triangles represent the PS1 colors of
unpublished spectroscopically confirmed $z\sim 6$ quasars (S.J. Warren et al., in preparation). Three of 
them were part of our candidate list (see Tables \ref{table:qsos_other} and \ref{table:qso_no_sel}).
The red diamonds are the new quasars presented in this paper (see Table \ref{table:ps1qso_stackprop}).
They are labeled with the following letters: $a=$  \Qa\ ($z=6.00)$, $b=$ \Qb\ ($z=5.99$),
$c=$ \Qc\ ($z=5.99$), $d=$ \Qd\ ($z=5.89$), $e=$ \Qe\ ($z=5.88$), $f=$ \Qf\ ($z=5.86$), 
$g=$ \Qg\ ($z=5.84$), $h=$ \Qeric\ ($z=5.73$), and \mbox{$i=$ \Qh\ ($z=5.70$)}.
}
\label{fig:selection}
\end{figure*}

We limited our survey to 
candidates with $\zps > 18.0$, S/N $>10$ in the \zps\ band, and S/N $>5$ in the \yps\
band (Equations (\ref{eq:zlimit}), (\ref{eq:zsn}), and (\ref{eq:ysn})).  We treated objects that were or not detected in the 
\ips\ band differently as summarized in Equation (\ref{eq:izcolor}). For the former we required a color $\ips - \zps > 2.2$,
for the latter we were less conservative and 
required $\ipsl - \zps > 2.0$, where \ipsl\ is the $3\sigma$-limiting magnitude. One of
our main contaminants are cool stars, which have similar $i-z$
colors to those of $z\gtrsim 5.7$ quasars but are redder in $z - y$. 
However, for quasars at $z\gtrsim 6.2$, a significant amount of flux in the $z$-band is
also absorbed, making these objects redder in the $z-y$ color (see Figure \ref{fig:selection}).
In this paper we aim to get the bulk of
the $z \sim 6$ quasar population while at the same time reducing the contamination by
L- and T-dwarfs. We produce a relatively pure sample by setting an upper limit in 
the $\zps - \yps$ color of $0.5$ (see Equation (\ref{eq:zycolor}) and Figure \ref{fig:selection}). 
This constraint has the drawback of preventing us from finding typical quasars at $z\gtrsim 6.2$. Once we have
a better understanding of our contaminants, we will relax this constraint to search for
quasars at higher redshifts. In order to prevent contamination from low redshift interlopers we also 
put constraints in the \rps\ and \gps\ bands.  In Equation (\ref{eq:rzcolor}), we will require that 
either the object was detected with a S/N$<3$ in the \rps\ band, or that the flux in the band was negative 
(i.e., the object was not detected or there was no coverage), or that there was a break in color of $\rps -\zps > 2.2$.
We required a non-detection or no coverage in the \gps\ band (Equation (\ref{eq:gcolor})), since
we expect virtually null flux in this band from a $z\sim 6$ quasar.
Quasars at these redshifts are expected to be unresolved.
In an attempt to select point-like sources we require the aperture magnitudes
and PSF magnitudes of our candidates to be consistent with each other within 0.3 mag at least in one of the detection bands 
(\zps\ or \yps, see\ Equation (\ref{eq:pointsource})). 

We can summarize the catalog selection criteria as follows:

\begin{subequations}
\begin{eqnarray}
\zps > 18.0& \label{eq:zlimit}\\
\mbox{S/N}(\zps) >  10& \label{eq:zsn} \\
\mbox{S/N}(\yps) > 5 & \label{eq:ysn} \\
((\ips < \ipsl) \; \mbox{AND} \, (\ips - \zps > 2.2)) \; \mbox{OR} & \nonumber \\
((\ips > \ipsl) \;  \mbox{AND}\, (\ipsl - \zps > 2.0)) & \label{eq:izcolor} \\
\zps - \yps < 0.5 & \label{eq:zycolor} \\
(\mbox{S/N}(\rps) < 3) \; \mbox{OR} \; (\mbox{flux}(\rps) \leq 0) \; \mbox{OR} &\nonumber \\
(\rps - \zps > 2.2) & \label{eq:rzcolor}\\
(\mbox{S/N}(\gps) < 3) \; \mbox{OR} \; (\mbox{flux}(\gps) \leq 0) & \label{eq:gcolor} \\
(-0.3 < \zpsa - \zps < 0.3) \; \mbox{OR} &\nonumber\\ 
(-0.3 < \ypsa - \yps <0.3) & \label{eq:pointsource} 
\end{eqnarray}
\end{subequations}

\subsection{Forced Photometry on Stacked Images} \label{sec:stack_phot}

In order to confirm and check the PS1 catalog colors, we implemented an algorithm to compute aperture photometry on the stacked images.
This algorithm retrieves 
$5\arcmin \times 5 \arcmin$ stacked postage stamps images centered on the candidates and their respective stacked catalogs. Then, we computed the aperture that maximizes the S/N
of bright stars in the field and used it to get the aperture photometry of the respective candidate in the image. Since the aperture photometry
is noisier than the PSF photometry, we relaxed the color criteria and required Equations (\ref{eq:zlimit}), (\ref{eq:izcolor}), (\ref{eq:zycolor}), and (\ref{eq:rzcolor})
to be consistent within $2\sigma$ with our own measurements from the stacked images. This step mainly got rid of candidates that had bad $\ips$\ photometry in 
the catalog. This is not surprising since our candidates were very faint or not detected in the $\ips$ band, and large photometric errors are expected.

We found empirically that when the diameter of our computed optimal apertures ($ \mbox{aperture\_diameter}$)
were smaller than  $0\sec75$ or 
greater than $3\sec 75$, the images had evident problems mostly related to background subtraction, 
causing unreliable catalog magnitudes. Thus, we 
only considered candidates with $0\sec 75 < \mbox{aperture\_diameter}(izy) < 3\sec 75$.

\subsection{Single Epoch Forced Photometry} \label{sec:single_phot}
It is possible that a fraction of our candidates were moving objects, spurious objects, or artifacts that appeared as very bright
sources in some of the individual exposures and did not appear in others. After stacking the images, such objects could still 
look like reasonable quasar candidates. For this reason, we performed forced aperture photometry on all \emph{individual} single epoch $z$ band images
(the detection band with the highest S/N) by using the optimal aperture determined in Section \ref{sec:stack_phot}. We flagged the images 
in which the S/N at the position of the candidate was S/N $< 3$ while, based on the magnitude of the candidate measured on the stacked image,
a S/N $>5$ was expected. We retained candidates that had less than 40\% flagged images and that the standard deviation of the single epoch magnitudes
was less than 1 mag. These criteria were determined based on tests performed with known quasars and objects that had a counterpart in SDSS and in the \textit{WISE} \citep[][]{wri10}.

\subsection{Cross-match with other Surveys} \label{sec:cross_surveys}
We required our candidates to be either undetected in 2MASS or, similarly to \cite{fan01},
that $\zps - \Jtmass < 1.5$.
We did forced photometry in the SDSS $i$-band (\isdss) and UKIDSS $J$-band (\Jukidss) when 
our candidates were in the area covered by these surveys. 
Candidates with S/N $< 5$ or that had S/N $>5$ in the \isdss\ band and that satisfy $\isdss - \zps > 2.0$ within $2\sigma$ errors were kept.
We also retained objects that had $\yps - \Jukidss < 1.5$.

\subsection{Cross-match with Known Quasars}
\label{sec:cross-match}
In order to avoid duplication with published quasars and other ongoing quasar search efforts, we removed fifteen published (including \Qeric)
and three unpublished quasars (ULASJ1207+0630 at $z=6.04$, ULASJ0148+0600 at $z=5.96$, and ULASJ1243+2529 at $z=5.83$; S.J.~Warren et al., in preparation, but
see also Mortlock et al. 2012) 
that were part of our candidate list. 
These quasars are discussed further in Section  \ref{sec:known_qsos}. All the remaining candidates were visually inspected.

\subsection{Imaging Follow-up} \label{sec:img_follow}

Obtaining deeper optical and NIR photometry is essential to confirm the reality and colors of our candidates and 
to remove cool dwarfs with similar optical colors to quasars or contaminants that could have scattered into our color selection.
The photometric follow-up observations were carried out over different observing runs and different instruments.
We used GROND \citep{gre08}, a $grizJHK$ simultaneous imager at the 2.2 m telescope in La Silla during 
2012 May (when a small region of the stacked catalog was available) and 2013 January, typical on-source exposure times were 1440 s in the 
NIR and 1380 s in the optical.
The ESO Faint Object Spectrograph and Camera 2  \citep[EFOSC2;][]{buz84} at the ESO New Technology Telescope (NTT) was used to perform 
imaging in the $i\# 705$ (\intt) and $z\#623$ (\zntt) bands during 2013 March with on-source exposure times  of 300 s. 
We used the Omega2000 camera \citep{biz98} at the 3.5m telescope in Calar Alto 
to perform imaging with the $z$ (\zotk), $Y$ (\Yotk), and $J$ (\Jotk) filters during 2013 March--June with 300 s exposure time on-source.
The data reduction was carried out using standard reduction steps, consisting of bias subtraction, flat fielding, sky subtraction, image alignment,
and finally stacking. 
The photometric zero points of the stacked images were calculated by using at least 5 stellar sources 
($\left| \mbox{mag}\ensuremath{_{\rm P1-aper}} - \mbox{mag}\ensuremath{_{\rm P1}} \right| < 0.2$) in the images with magnitudes and colors from the PV1 catalog.
Since all the filter curves are not exactly the same, conversions between PS1 and GROND/EFOSC2/Omega2000 magnitudes were computed as follows:
first, we computed spectral synthetic magnitudes of 
template stars of various spectral classes (O -- K) in each photometric system. We then produced color--color diagrams for various filters 
(e.g., $i_{\rm GROND}-\ips$ vs $\ips-\zps$). Through these plots we have inferred the photometric system correction terms via linear fits 
of the stellar loci. 
The color conversions  are the following:
\begin{subequations}
\begin{eqnarray*}
\igrond = & \ips - 0.089 \times (\rps - \ips) + 0.001  \label{eq:PS_iGROND}\\
\zgrond = &\zps - 0.214 \times (\zps - \yps)   \label{eq:PS_iGROND}\\
\Jgrond = &\Jtmass - 0.012 \times (\Jtmass - \Htmass) + 0.004 \label{eq:PS_JGROND}\\
\Hgrond = &\Htmass + 0.030 \times (\Htmass - \Ktmass) + 0.009 \label{eq:PS_HGROND}\\
\intt = &\ips - 0.149 \times (\ips - \zps) - 0.001  \label{eq:PS_iNTT}\\
\zntt = &\zps - 0.265 \times (\zps - \yps)  \label{eq:PS_zNTT}\\
\zotk = & \zps - 0.245 \times (\zps - \yps) \\
\Yotk = & \yps - 0.413 \times (\zps - \yps) +0.012\\
\Jotk = & \Jtmass + 0.093  \times (\Jtmass - \Htmass)
\end{eqnarray*}
\end{subequations}

\noindent where \Jtmass, \Htmass, and \Ktmass\ are 2MASS magnitudes in the AB system.
The accuracy of the photometric zero points was typically 0.01 -- 0.04 mag and are included in the magnitude errors presented in this work.

\subsection{Spectroscopic Follow-up} \label{sec:spc_follow}
In order to confirm the nature and redshifts of the candidates, we have carried out optical and NIR spectroscopy using EFOSC2
at the NTT telescope in La Silla, the FOcal Reducer/low dispersion Spectrograph 2 \citep[FORS2;][]{app92}
at the Very Large Telescope (VLT), the Multi-Object Double Spectrograph  \citep[MODS;][]{pog10} at the Large Binocular Telescope (LBT), 
and the Folded-port InfraRed Echellette \citep[FIRE;][]{sim08, sim13} spectrometer in the Baade Telescope at Las Campanas Observatory.
The details of the spectroscopy observations of the new PS1 quasars are shown in Table \ref{table:spc_obs}.
The data reduction included bias subtraction, flat fielding using lamp flats and sky subtraction. 
For the wavelength calibration exposures of He, HgCd and Ne arc lamps were obtained. The wavelength calibration was checked 
using sky emission lines. The typical rms of the wavelength calibration was better than 0.5 \AA.
For the flux calibrations we used observations of the spectrophotometric standard stars
G158-100, GD50, LTT7379, HD49798, EG274, and G191-B2B \citep{oke90,ham92,ham94}.

\begin{table*}[htdp]
\caption{Spectroscopic Observations of the New PS1 Quasars}
\begin{center}
\begin{tabular}{lccccc}
\hline
\hline
QSO		      & Instrument	&	Date  		& Slit Width (arcsec)	 & Exp. Time (s) & Seeing (arcsec)	 \\
\hline
PSO J340.2041--18.6621 &	 EFOSC2		&	2012 Jun 21	& 1.5		         & 	3600     & 0.80 -- 1.40    \\
		      &  MODS		&	2012 Nov 17	& 1.2		         & 	3000     & 0.90 -- 1.30     \\
PSO J007.0273+04.9571 & FORS2		&	2013 Jul 9	& 1.3			 &      1782	 & 0.75 -- 1.01       \\
PSO J037.9706--28.8389 & FORS2		& 	2013 Mar 4-5	& 1.3			 &	3600	 & 0.66 -- 0.90     \\
PSO J187.3050+04.3243 & FORS2		&	2013 Apr 12	& 1.3			 & 	2682	 & 1.05 -- 1.17      \\
PSO J213.3629--22.5617 & FORS2		&	2013 May 3	& 1.3			 & 	1782	 & 0.56 -- 0.61       \\
PSO J183.2991--12.7676 & FORS2		&	2013 Apr 13	& 1.3			 & 	1782	 & 0.62 -- 0.77      \\
		       & FIRE		&	2013 Apr 19	& 0.6			 & 	6000	 & 0.67  -- 1.22   \\
PSO J210.8722--12.0094 & FORS2		&	2013 May 9	& 1.3		         &      2682 	 & 0.66 -- 0.71      \\
PSO J045.1840--22.5408 & FORS2		&	2013 Aug 9	& 1.3         		& 	1782	 & 0.82 -- 0.90      \\

\hline
\end{tabular}
\end{center}
\label{table:spc_obs}
\end{table*}

\section{Eight new quasars at redshift $\sim 6$}

\subsection{Spectra}
We have taken spectra of nine objects that were photometrically followed up. Eight of them satisfied our criteria in Section \ref{ps1_selection}
and they all turned out to be quasars at $z\sim 6$. One object, selected with preliminary criteria, was found to be an M-star, 
however this source did not satisfy the final color cuts presented in this paper.
None of the eight quasars is detected in the NRAO VLA Sky Survey \citep[][]{con98}. Only \Qb\ and \Qd\ are located in the region covered by 
the Faint Images of the Radio Sky at Twenty cm (FIRST) radio survey \citep{bec95} but they are undetected. Four quasars, \Qa, \Qb, \Qc, and \Qf, 
are in the \textit{WISE} All-Sky data release products catalog (Cutri et al. 2012) with $>3.0\sigma$ detections.  The \textit{WISE} magnitudes are tabulated in Table
\ref{table:wise} in Appendix \ref{ap:wise}.

The PS1 stacked catalog and the follow-up photometry of the quasars are presented in Tables \ref{table:ps1qso_stackprop} 
and \ref{table:ps1qso_followup}. The optical spectra of the new quasars are shown in Figure \ref{fig:spectra}.
Each spectrum has been scaled to the corresponding \zps\ magnitude. For completeness, the  
first PS1 quasar \Qeric\ \citep{morg12}
is also included in Figure \ref{fig:spectra} and Table \ref{table:ps1qso_stackprop}.

\subsection{Redshift Determination}
\begin{figure*}[htbp]
\begin{center}
 \centerline{\includegraphics[width=.9\linewidth]{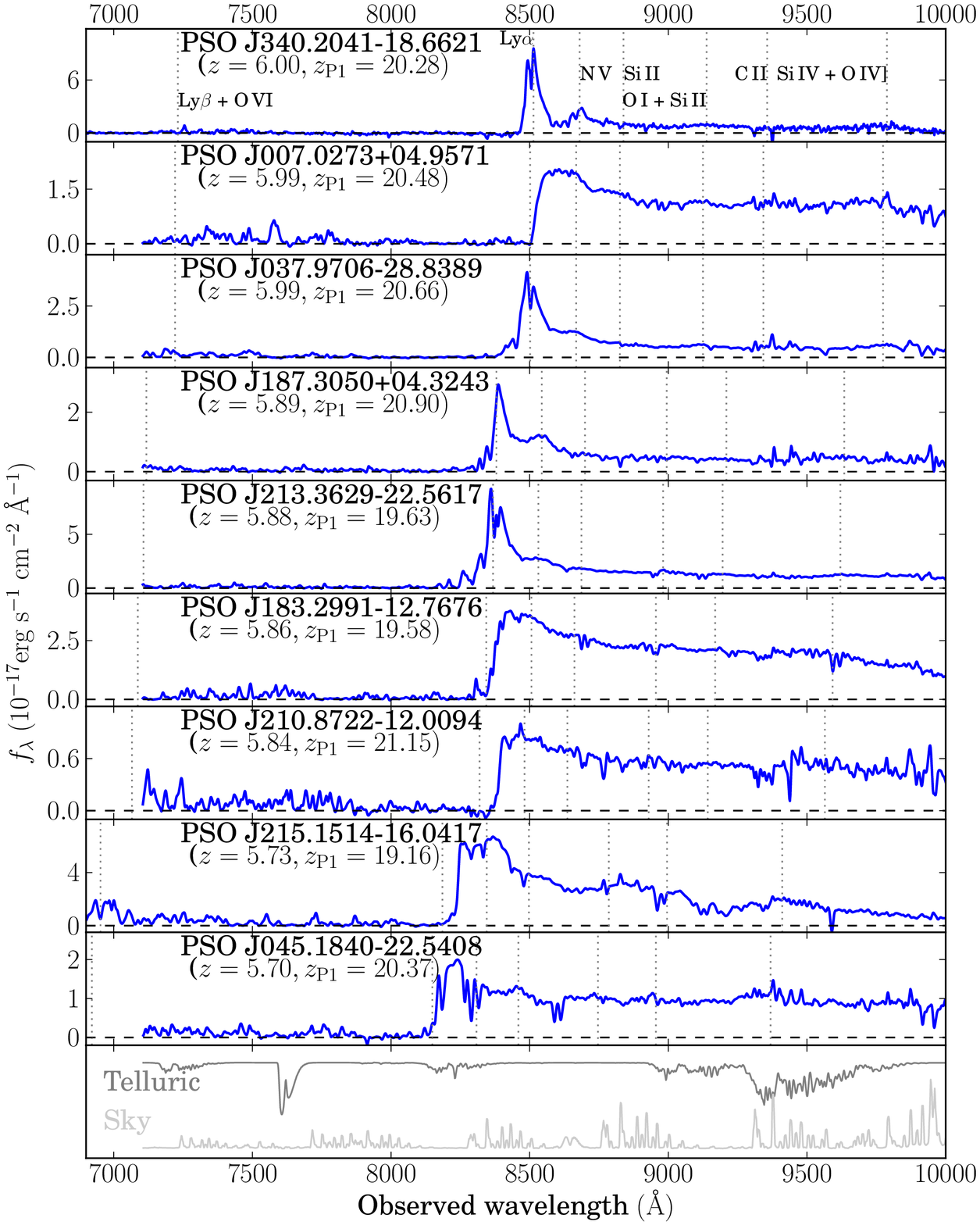}}
 \caption{Spectra of the nine newly discovered PS1 quasars at $5.70\leq z \leq 6.00$. The quasar \Qeric\ was published in \cite{morg12}.
 Here we show the discovery spectra of the new quasars (with the exception of \Qa\ whose
 EFOSC2 discovery spectrum had very low S/N). The bottom panel shows the median telluric absorption and sky emission lines.
 Vertical dashed lines indicate the observed wavelengths of key spectral lines, as given in the top panel.
 \label{fig:spectra}}
 \end{center}
 \end{figure*}

There are four quasars, \Qa, \Qc, \Qd, and \Qe\ that show clear emission lines and their redshift estimation 
was performed by Gaussian fitting to the N$\,${\sc v}, $\lambda 1240$ (hereafter \nv),
\oi\ $\lambda 1305$ (hereafter \oi), and/or \siivpoiv\ $\lambda 1398 $ (hereafter \siivpoiv) lines.
The observed wavelengths in the composite spectra from \cite{van01} 
were taken as reference.  The uncertainties in our line fittings were negligible compared to 
the intrinsic shifts known to exist with respect to the quasar systemic redshift \citep{ric02a, she07}. Thus, following previous studies
\citep[e.g.,][]{fan06a, jia09, will09}, we  adopted redshift uncertainties of $\Delta \,z = 0.02$. The other four quasars, \Qb, \Qf, \Qg, and \Qh\ have
discovery spectra with a bright continuum almost devoid of bright emission lines (see Figure \ref{fig:spectra}). In the absence 
of strong lines, an accurate redshift estimate for these quasars is challenging. We estimated their redshift by matching their continuum to the composite spectra from \cite{van01} and \cite{fan06a}. 
For the brightest of these quasars, \Qf, we obtained a second spectrum with a higher S/N and extended wavelength 
coverage from which we were able to 
estimate its redshift by fitting weak emission lines. The redshift estimated by the Gaussian fitting was off by 0.03 with respect to our
preliminary estimation based on the continuum matching (for more details see Section \ref{sec:qf}). Based on the case of \Qf, we  adopted a redshift 
uncertainty of $\Delta \,z = 0.05$ for the quasars whose redshifts were determined only by matching their continuum to templates.
Once the redshifts were estimated, we calculated the absolute magnitude of the continuum rest frame 1450 \AA\ ($M_{1450}$) for each quasar,
by fitting a power law of the form $f_\lambda = C \times \lambda^\beta$ to regions of the continuum that are generally uncontaminated by
emission lines (1285--1295, 1315--1325, 1340--1375, 1425--1470, 1680--1710, 1975--2050, and 2150--2250 \AA) and that were not affected by
significant errors or absorption features. 
The $M_{1450}$ values are listed in Table \ref{table:ps1qso_stackprop}.

\subsection{Notes on Individual Objects}

We here discuss the individual quasars, sorted by descending redshift.

 \subsubsection{\Qa\ ($z=6.00 \pm 0.02$)} 
 This quasar was selected from a preliminary small region of the PS1 stacked catalog.
Because of the low S/N of its EFOSC2 discovery spectrum,  it was difficult to determine an accurate redshift.
 Thus, we have obtained a higher S/N spectrum using
 MODS at the LBT (see Table \ref{table:spc_obs}). The MODS spectrum is shown
 in Figure \ref{fig:spectra}. The spectrum shows a strikingly strong and narrow \lya\ emission line. A separate narrow
 \nv\ line is evident and the \oi\ line is also detected. 
 The redshift was calculated from the average of the Gaussian fits to 
 the \nv\ and \oi\ lines. We have also carried out 870~$\mu$m observations of \Qa, however the quasar remained 
 undetected in the map with an rms of 1.6 mJy
 at the quasar position.  Details of the submillimeter observations can be found in Appendix \ref{ap:submm}.

\subsubsection{\Qb\ ($z=5.99 \pm 0.05$)}
The discovery spectrum shows a weak-line emission quasar and with the 
current S/N there are not clear emission lines from which
the redshift could be determined.
The redshift estimate was calculated by fitting the continuum to the composite spectra from \cite{van01} and \cite{fan06a}.
This quasar was independently discovered by L. Jiang et al. (in preparation).

\subsubsection{\Qc\ ($z=5.99 \pm 0.02$)}
The discovery spectrum shows a bright \lya\ line and a well separated \nv\ line. The  \oi\ emission line is also detected. 
The redshift was estimated as the average of the best-fit values of the \nv\ and \oi\ lines.

\subsubsection{\Qd\ ($z=5.89 \pm 0.02$)}
Bright and narrow \lya\ and \nv\ emission lines are clear in the discovery spectrum. 
There is a tentative  \siivpoiv\
emission line but at very low S/N
and in a region with considerable telluric absorption. The redshift was calculated from the Gaussian fitting to the \nv\ emission line.

\subsubsection{\Qe\ ($z=5.88 \pm 0.02$)}
Bright \lya\ and \nv\ emission lines are evident in the discovery spectrum. The \siivpoiv\  and \oi\ emission lines are 
also detected. The redshift estimation was calculated from the best-fit  values of the \nv\ and \siivpoiv\
lines. Even though adding the fit of the \oi\ line to the redshift determination did not change the
estimated value, we decided to not consider it since it seemed to be affected 
by an absorption system. We identify a possible absorption system at $z=4.780 \pm 0.002$ from the \civ\ $\lambda\lambda 1548, 1550$
absorption feature at $\lambda = (8945.9\, \text{\AA}, 8960.2 \, \text{\AA})$. However, we note that the scatter between the 
expected and observed positions of the two lines is large ($\sim 1$ \AA). The spectrum shows also a clear \mgii\ $\lambda\lambda 2796, 2803$
absorption doublet
at $\lambda = (9113.0\, \text{\AA}, 9136.3 \, \text{\AA})$, corresponding to an absorber at $z=2.2594 \pm 0.0005$. This absorber is further confirmed
by the presence at the same redshift of the \feii\ $\lambda 2586$ absorption at $\lambda=8429.8$ and
\feii\ $\lambda 2600$ \AA\ absorption at $\lambda=8474.0$ \AA.

\subsubsection{\Qf\ ($z=5.86 \pm 0.02$)} \label{sec:qf}

\begin{figure}
\epsscale{1.0}
\plotone{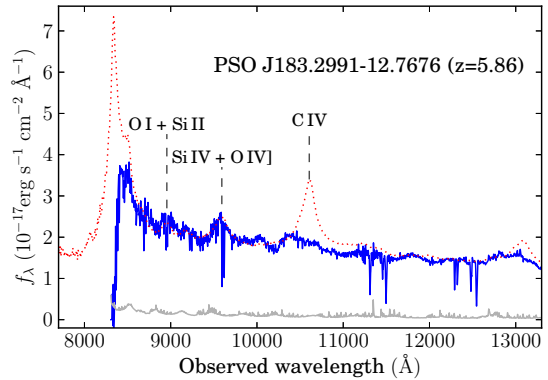}
\caption{FIRE spectrum of the quasar \Qf\ ($z=5.86$). The gray solid line around zero flux shows the $1\sigma$ error. The dotted line shows the composite spectrum
from \cite{van01} as reference. Unlike the discovery spectrum in Figure \ref{fig:spectra}, the \oi\ and \siivpoiv\ emission lines are clearly identified. \label{fig:fire_spc}}
\end{figure}

This is the brightest quasar in our new sample (but fainter than \Qeric\ reported in Morganson et al. 2012).
Figure \ref{fig:spectra} shows the FORS2 discovery spectrum. This spectrum has a very bright continuum but it does not show any
detectable emission line and it seems that \lya\ is almost completely absorbed. Since we did not have emission lines to fit,
we estimated a preliminary redshift from fitting the 
continuum to the composite spectra from \cite{van01} and \cite{fan06a}, yielding $z=5.83$. We obtained a second spectrum of higher S/N of this object
using the FIRE spectrometer (see Table \ref{table:spc_obs}). 
The FIRE spectrum was scaled to the \yps\ magnitude and is presented in Figure \ref{fig:fire_spc}. The spectrum  shows clear \oi\ and \siivpoiv\ emission lines which were used for the final redshift estimation.
The \oi\ and \siivpoiv\ lines are affected by absorption systems, but the signal is high enough to get good fits even when masking the absorbed regions. 
There is a tentative \cii\ emission line at lower S/N also affected by absorption that was not used for the redshift estimate.
In the case of this quasar, the redshift estimated by matching the continuum
was off by 0.03 with respect to the line fitting estimate ($z = 5.86$). Based on this case, we assumed a redshift uncertainty of $0.05$ for the other weak-line emission quasars in
our sample whose redshift estimate was solely based on continuum fitting. 
We identify absorptions systems at $z=4.460 \pm 0.002$, $z=2.1074 \pm 0.0005$, and $z=2.431 \pm 0.001$ from the
\civ\ $\lambda\lambda 1548, 1550$ doublet at $\lambda = (8946.2\, \text{\AA}, 8961.3\, \text{\AA})$, \mgii\ $\lambda\lambda 2796, 2803$ doublet at $\lambda = (8688.5\, \text{\AA}, 8709.8\, \text{\AA})$,
and another \mgii\ $\lambda\lambda 2796, 2803$ doublet at $\lambda = (9593.4\, \mbox{\AA}, 9618.6\, \text{\AA})$ respectively. For the latter we also identify
the associated \feii\ $\lambda 2586$ absorption at $\lambda=8874.8$ and
the  \feii\ $\lambda 2600$ \AA\ absorption at $\lambda=8921.3$ \AA. There seems to be a broad absorption line blueshifted from the expected location
of the \civ\ $\lambda 1546$ (hereafter \civ) emission line, which is remarkably absorbed in the spectrum of this quasar.

\subsubsection{\Qg\ ($z=5.84 \pm 0.05$)}

This is the faintest of the PS1 quasar sample ($\zps=21.15\pm0.08$). The discovery spectrum does not have a S/N high enough
to identify clear features in the quasar, it seems rather a weak-line emission or lineless quasar. The redshift estimate 
had to be determined
by fitting the continuum to the composite spectra of \cite{van01} and \cite{fan06a}. There seems to be a weak absorption feature at
$\lambda\lambda = 8692.6, 8705.3$. If real, it could be associated with a \civ\ $\lambda\lambda 1548, 1550$ absorption system
at $z=4.616 \pm 0.001$, although with
the current S/N it is hard to tell.

\subsubsection{\Qh\ ($z=5.70 \pm 0.05$)}
This is the lowest redshift quasar in the PS1 sample so far. 
The discovery spectrum of this quasar looks very peculiar and seems to be significantly affected by absorption systems or intrinsic absorption 
from the host galaxy. The spectrum shows a weak-line emission quasar and the S/N does not allow us to identify emission lines in regions
not affected by significant telluric absorption or sky emission to calculate the redshift from. The redshift was
estimated by matching the continuum to the composite spectra from \cite{van01} and \cite{fan06a}. There are tentative \SIii\ and \siivpoiv\ emission lines.
If real, the redshift estimated from the average of their Gaussian fits correspond to $z=5.69$, in agreement with our continuum fitting. 
However, we conservatively estimated the redshift uncertainty as $\Delta \, z=0.05$ as for the other 
weak emission line quasars where the redshift was calculated only from continuum matching.
There are also several absorption lines around $8300$ \AA\ (e.g., $\lambda=8265.2,\, 8281.3,\, 8306.8$, and $8315.6$ \AA).
Most of the lines are blended and, given the resolution of the spectrum, it is hard to unambiguously identify them. These lines could be due to a 
group of galaxies at $z\sim 1.96$ if we assume that all the lines are due to \mgii\ absorbers. We identify a \mgii\ $\lambda\lambda 2796, 2803$ doublet at 
$z=2.0721 \pm 0.0005$ (observed at $\lambda=8589.2$ \AA\ and $8611.9$ \AA). The corresponding \mgi\ $\lambda 2853$ \AA\ seems marginally 
detected at $\lambda= 8764.0$ \AA.

\begin{table*}[htdp]
\caption{PS1 Stacked Catalog Magnitudes, Redshifts, and Absolute Magnitudes of the PS1 High-redshift Quasars.}
\begin{center}
\begin{tabular}{lccccccc}
\hline
\hline
QSO		      &	R.A. (J2000)	&	Decl. (J2000)  & \ips \tablenotemark{a}		 & \zps		     & \yps 		  &  Redshift & $M_{1450}$  \\
\hline
PSO J340.2041--18.6621 &	 22:40:48.98	&	--18:39:43.8	& $>23.44$ 	  & $20.28 \pm 0.05$  & $20.47 \pm 0.13$ & 6.00  &  $-26.0$   \\
PSO J007.0273+04.9571 & 00:28:06.56	&	+04:57:25.7	&$23.24 \pm 0.29$ & $20.48 \pm 0.06$  & $20.18 \pm 0.08$ & 5.99  &  $-26.5$  \\
PSO J037.9706--28.8389 & 02:31:52.96	& 	--28:50:20.1	&$>22.76$   	  & $20.66 \pm 0.09$  & $20.70 \pm 0.21$ & 5.99   &  $-25.6$ \\
PSO J187.3050+04.3243 & 12:29:13.21	&	+04:19:27.7	&$23.19 \pm 0.28$ & $20.90 \pm 0.05$  & $21.09 \pm 0.18$ & 5.89  & $-25.4$    \\
PSO J213.3629--22.5617 & 14:13:27.12	&	--22:33:42.3	&$22.98 \pm 0.34$ & $19.63 \pm 0.04$  & $19.83 \pm 0.11$ & 5.88   & $-26.6$   \\
PSO J183.2991--12.7676 & 12:13:11.81	&	--12:46:03.5	&$22.32 \pm 0.15$ & $19.58 \pm 0.04$  & $19.31 \pm 0.05$ & 5.86   &  $-27.3$ \\
PSO J210.8722--12.0094 & 14:03:29.33	&	--12:00:34.1	&$>23.17$         & $21.15 \pm 0.08$  & $20.69 \pm 0.13$ & 5.84   &  $-25.7$  \\
PSO J215.1514--16.0417\tablenotemark{b} &	14:20:36.34	& --16:02:30.2 & $21.75\pm 0.09$  & $19.16 \pm 0.02$  & $19.17 \pm 0.08$ & 5.73 & $-27.6$\tablenotemark{c}	\\
PSO J045.1840--22.5408 & 03:00:44.18	&	--22:32:27.2	& $>22.79$         & $20.37 \pm 0.06$  & $20.53 \pm 0.15$ & 5.70 &   $-26.4$   \\

\hline
\end{tabular}
\tablenotetext{1}{The lower limits correspond to $3\sigma$ limiting magnitudes.}

\tablenotetext{2}{Quasar published in \cite{morg12}, here we have slightly modified the coordinates from PSO J215.1512-16.0417 to 
\mbox{PSO J215.1514--16.0417} based on the stacked catalog.}

\tablenotetext{3}{To calculate this absolute magnitude, we took the best continuum fit from \cite{morg12} and scaled it to the \zps\ stack magnitude presented here.}
\end{center}
\label{table:ps1qso_stackprop}
\end{table*}

\begin{table*}[htdp]
\caption{Follow-up Photometry of the New PS1 Quasars.}
\begin{center}
\begin{tabular}{lcccccc}
\hline
\hline
QSO		      &	\igrond		 &	\zgrond    & \Jgrond	       & \Hgrond	  & \intt	      &  \zntt  \\
\hline
PSO J340.2041--18.6621 &	 $23.32 \pm 0.17$&$20.11 \pm 0.04$& $20.28 \pm 0.08$ & $19.9 \pm 0.11$  & $\cdots$       & $\cdots$     \\
PSO J007.0273+04.9571 & $22.74 \pm 0.24$&$20.39 \pm 0.04$& $19.77 \pm 0.11$ & $19.71 \pm 0.15$ & $\cdots$ 		      & $\cdots$         \\
PSO J037.9706--28.8389 & $23.19 \pm 0.15$&$20.78 \pm 0.04$& $20.41 \pm 0.14$ & $20.59 \pm 0.26$ & $\cdots$ 		      & $\cdots$      \\
PSO J187.3050+04.3243 & $\cdots$	 &$\cdots$          &	$\cdots$	      &	 $\cdots$	  & $21.85 \pm 0.04$ & $21.00 \pm 0.04$      \\
PSO J213.3629--22.5617 &$\cdots$	 & $\cdots$          &	$\cdots$	      &	$\cdots$	  & $20.70 \pm 0.02$ & $19.77 \pm 0.01$       \\
PSO J183.2991--12.7676 &$\cdots$		 & $\cdots$          &	$\cdots$	      &	 $\cdots$	  & $20.69 \pm 0.03$ & $19.31 \pm 0.02$      \\
PSO J210.8722--12.0094 & $\cdots$		 & $\cdots$	          &	$\cdots$	      &	 $\cdots$	  & $22.35 \pm 0.06$ & $21.00 \pm 0.03$    \\
PSO J045.1840--22.5408 & $22.78 \pm 0.10$&$20.35 \pm 0.03$& $19.65 \pm 0.08$ & $19.42 \pm 0.08$ & $\cdots$& $\cdots$      \\

\hline
\end{tabular}
\end{center}
\label{table:ps1qso_followup}
\end{table*}

\begin{table*}[htdp]
\caption{High-Redshift Quasars from other Surveys that were Part of the Final Candidate List in this Work.}
\begin{center}
\begin{tabular}{cccccccc}
\hline
\hline
QSO 		 &	R.A. (J2000)\tablenotemark{a}	&	Decl. (J2000)\tablenotemark{a}  &  \ips\tablenotemark{b}	     & \zps 		    & \yps		&  Redshift 	& Reference \\
\hline
SDSSJ1030+0524  &	10:30:27.12	&	+05:24:55.1	& $>22.98$ 	     & $20.10 \pm 0.05$  & $20.28 \pm 0.20$   & 6.28	    	& 1 \\
SDSSJ1623+3112  & 	16:23:31.81	&	+31:12:00.5	&$>23.27$ 	     & $20.20 \pm 0.06$  & $20.22 \pm 0.10$   & 6.22 		& 1 \\
SDSSJ1250+3130  & 	12:50:51.91	&	+31:30:21.8	&$23.55 \pm 0.39$  & $19.94 \pm 0.04$  & $20.26 \pm 0.12$   & 6.13 		& 1 \\
SDSSJ1602+4228  & 	16:02:53.95	&	+42:28:25.0	&$22.56 \pm 0.20$  & $20.09 \pm 0.03$  & $19.71 \pm 0.06$   & 6.07	 	& 1 \\
SDSSJ1630+4012  & 	16:30:33.90	&	+40:12:09.7	&$23.02 \pm 0.35$  & $20.37 \pm 0.07$  & $20.58 \pm 0.12$   & 6.05 		& 1 \\
ULASJ1207+0630  & 	12:07:37.44	&	+06:30:10.2	&$>23.22$	     & $20.44 \pm 0.04$  & $20.19 \pm 0.11$   & 6.04		& 2 \\
SDSSJ1137+3549  &	11:37:17.73	&	+35:49:56.9	&$22.15 \pm 0.11$  & $19.43 \pm 0.02$  & $19.44 \pm 0.05$   & 6.01 		& 1 \\
ULASJ0148+0600  & 	01:48:37.64	&	+06:00:20.1	&$22.80 \pm 0.25$  & $19.46 \pm 0.02$  & $19.40 \pm 0.04$   & 5.96 		& 2 \\
SDSSJ1335+3533  & 	13:35:50.81	&	+35:33:15.9	&$22.57 \pm 0.28$  & $20.27 \pm 0.03$  & $19.97 \pm 0.08$   & 5.95 		& 1 \\
SDSSJ1411+1217  & 	14:11:11.29	&	+12:17:37.3	&$23.25 \pm 0.26$  & $19.57 \pm 0.02$  & $20.08 \pm 0.08$   & 5.93		& 1 \\
SDSSJ0005--0006  & 	00:05:52.34	&	--00:06:55.7	&$23.09 \pm 0.20$  & $20.50 \pm 0.06$  & $20.69 \pm 0.14$   & 5.85	        & 1 \\
NDWFSJ1425+3254 & 	14:25:16.33	&	+32:54:09.5	&$23.40 \pm 0.37$  & $20.44 \pm 0.04$  & $20.33 \pm 0.11$   & 5.85	        & 3 \\
ULASJ1243+2529  & 	12:43:40.82	&	+25:29:23.8	&$>23.51$ 	     & $20.18 \pm 0.05$  & $20.61 \pm 0.14$   & 5.83		& 2 \\
SDSSJ1436+5007  & 	14:36:11.73	&	+50:07:07.2	&$22.57 \pm 0.23$  & $20.24\pm0.06$    & $20.24 \pm 0.08$   & 5.83 		& 1 \\
SDSSJ0002+2550  & 	00:02:39.39	&	+25:50:35.0	&$21.93 \pm 0.08$  & $19.02 \pm 0.05$  & $19.53 \pm 0.07$   & 5.80 		& 1 \\
SDSSJ1044--0125  & 	10:44:33.04	&	--01:25:02.1	&$21.87 \pm 0.12$  & $19.35 \pm 0.02$  & $19.29 \pm 0.07$   & 5.7847		& 4 \\
ULASJ0203+0012  &	02:03:32.38	&	+00:12:29.3	& $>23.85$ 	     & $20.78 \pm 0.09$  & $20.50 \pm 0.12$   & 5.72 		& 5 \\
\hline
\end{tabular}
\tablerefs{(1) \cite{fan06b}; (2) S.J.~Warren et al., in preparation; (3) \cite{coo06}; (4) \cite{wan13}; (5) \cite{mor09}  }
\tablenotetext{1}{The coordinates correspond to the coordinates in the PS1 stacked catalog and are not necessarily identical to the ones in the discovery papers.}
\tablenotetext{2}{The lower limits correspond to $3\sigma$ limiting magnitudes.}

\end{center}
\label{table:qsos_other}

\end{table*}

\begin{table*}[htdp]
\caption{High-redshift Quasars from other Surveys that did not Satisfy the Color Selection of this Work}
\begin{center}
\begin{tabular}{cccccccc}
\hline
\hline
QSO 		 &	R.A. (J2000)\tablenotemark{a}	&	Decl. (J2000)\tablenotemark{a}  &  \ips \tablenotemark{b}     	     & \zps 		    & \yps		&  Redshift 	& Reference \\
\hline
SDSSJ1148+5251  &	11:48:16.65	&	+52:51:50.4	& $>23.07$ 	     & $20.63 \pm 0.04$  & $19.42 \pm 0.10$   & 6.42	    	& 1 \\
ULASJ1148+0702  &	11:48:03.29	&	+07:02:08.3	& $>22.71$ 	     & $21.03 \pm 0.08$  & $20.44 \pm 0.15$   & 6.29 		& 2 \\
SDSSJ1048+4637  & 	10:48:45.07	&	+46:37:18.5	&$22.98 \pm 0.38$  & $20.19 \pm 0.04$  & $19.49 \pm 0.12$   & 6.20		& 1 \\
ULASJ1319+0950  & 	13:19:11.30	&	+09:50:51.5	&$22.56 \pm 0.19$  & $20.14 \pm 0.04$  & $19.53 \pm 0.07$   & 6.1330		& 3 \\
CFHQSJ1509--1749 &	15:09:41.78	&	--17:49:26.8	& $>22.06$ 	     & $20.25 \pm 0.07$  & $19.61 \pm 0.10$   & 6.12	 	& 4 \\
SDSSJ0353+0104  & 	03:53:49.73	&	+01:04:04.7	&$>23.12$	     & $21.15 \pm 0.10$  & $20.69 \pm 0.19$   & 6.049		& 5 \\
SDSSJ2054--0005  & 	20:54:06.50	&	--00:05:14.4	&$22.65 \pm 0.20$  & $21.01 \pm 0.09$  & $20.66 \pm 0.17$   & 6.0391		& 3 \\
SDSSJ2310+1855  &	23:10:38.89	&	+18:55:19.9	& $21.55 \pm 0.10$  & $19.64 \pm 0.04$  & $19.08 \pm 0.04$   & 6.0031		& 2 \\
SDSSJ0927+2001  & 	09:27:21.82	&	+20:01:23.5	&$21.52 \pm 0.16$  & $19.86 \pm 0.03$  & $19.88 \pm 0.11$   & 5.77 		& 6 \\
SDSSJ1621+5155  & 	16:21:00.94	&	+51:55:48.8	&$21.96 \pm 0.14$  & $20.05 \pm 0.06$  & $19.80 \pm 0.08$   & 5.71 		& 7 \\

\hline
\end{tabular}
\tablerefs{(1) \cite{fan06b}; (2) S.J~Warren et al., in preparation; (3) \cite{wan13}; (4) \cite{will07}; (5) \cite{jia08}; (6) \cite{car07}; (7) \cite{wan08}  }
\tablenotetext{1}{The coordinates correspond to the coordinates in the PS1 stacked catalog and are not necessarily identical to the ones in the discovery papers.}
\tablenotetext{1}{The lower limits correspond to $3\sigma$ limiting magnitudes.}

\end{center}
\label{table:qso_no_sel}
\end{table*}

\section{Known quasars in \PS}
\label{sec:known_qsos}
\begin{figure}
\epsscale{1.0}
\plotone{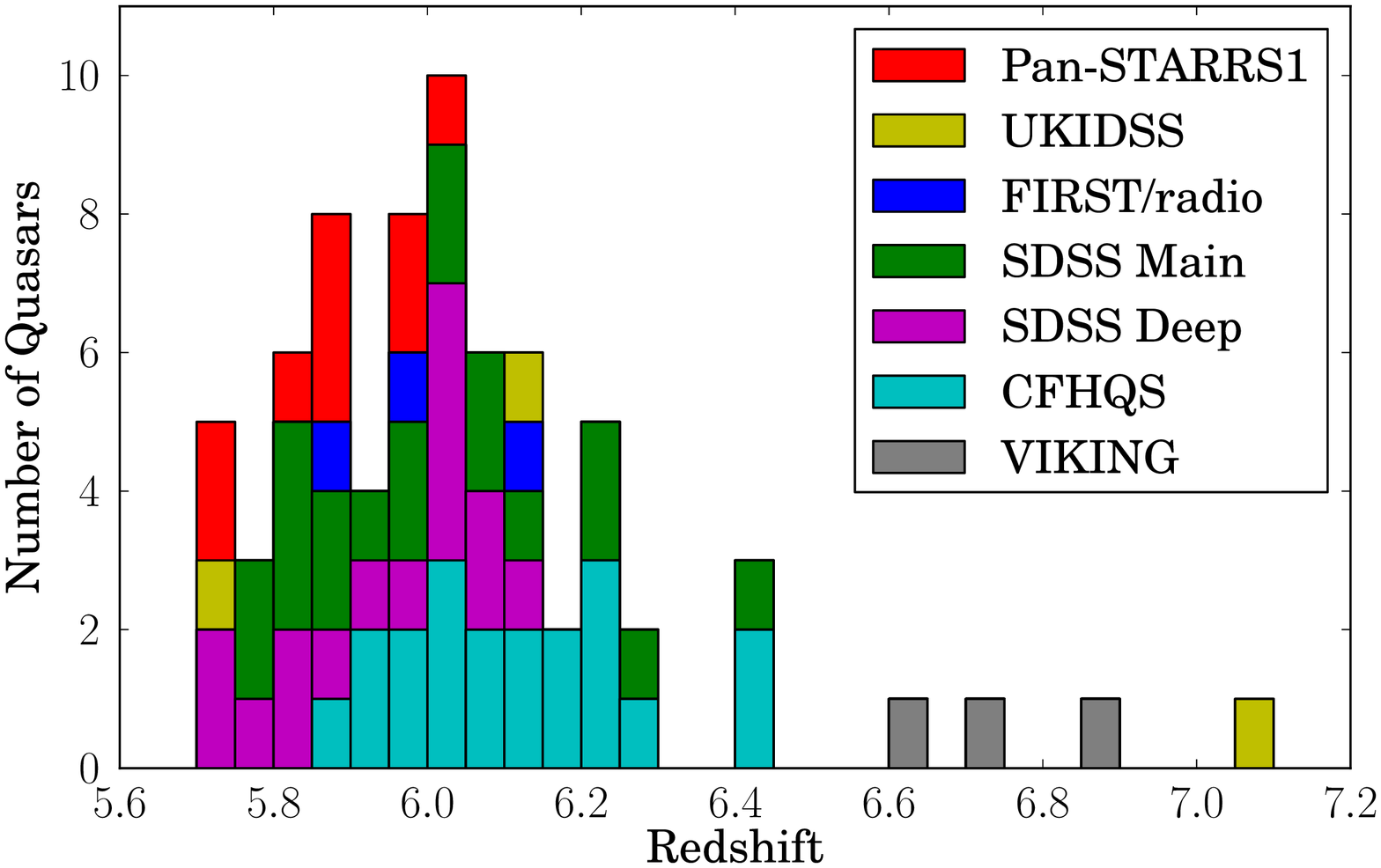}
\caption{Redshift distribution of the $z\geq 5.7$ quasars published to date. \PS\ includes the quasars in \cite{morg12} and this work. 
UKIDSS includes the quasars in \cite{ven07a} and \cite{mor09, mor11}. FIRST/radio includes the quasars in \cite{mcg06}, \cite{coo06}, and \cite{zei11}.
SDSS Main includes the quasars in \cite{fan01, fan03,fan04, fan06a}. SDSS Deep includes the quasars in 
\cite{mah05}, \cite{got06}, \cite{jia08}, \cite{wan08}, \cite{jia09},  \cite{der11}, and \cite{wan13}.
CFHQS includes the quasars in \cite{will07, will09, will10, will10b}. VIKING includes the quasars in \cite{ven13}.
}
\label{fig:qsos_hist}
\end{figure}

As mentioned in Section \ref{sec:cross-match}, the 17 known quasars in Table \ref{table:qsos_other} and \Qeric\ in Table \ref{table:ps1qso_stackprop}
were part of our candidate list. In order to calculate the fraction of known quasars that we recover using our current selection strategy, we
cross-matched a list of the known quasars at $z>5.7$ with our PS1 stacked catalog. We required the quasars to satisfy the same criteria as
in Section \ref{sec:selection} except for the magnitude and color cuts (Equations (\ref{eq:zlimit}), (\ref{eq:izcolor}), (\ref{eq:zycolor}),
(\ref{eq:rzcolor}), and (\ref{eq:gcolor})). In our current catalog, there are 36 known quasars that satisfy these criteria,
including the quasars discovered in this work.
Other quasars do not appear in our catalog for a variety of reasons
including (1) they are detected in PS1 but flagged or not detected in at least one of the
bands in which we are requiring detections (\ips, \zps, and \yps), 
(2) they are in regions without coverage i.e., chip gaps,
quasars with $92\deg<\,$R.A.$<132\deg$ (our catalog has no coverage at this R.A. range), or quasars with Decl. $< -30\deg$; (3) they are too faint for the current depth of PS1 
(i.e.,  S/N(\zps)$<10$ or S/N(\yps)$<5$). 
The PS1 colors of published quasars and the unpublished quasars from S.J.~Warren et al. (in preparation) are shown in  Figure \ref{fig:selection} with blue 
squares and green triangles respectively. Table \ref{table:qso_no_sel} shows the PS1 photometry of the known quasars that
are in the PS1 catalog but do not satisfy our color selection.
In total, we are recovering $72\%$ (26/36) of the high-redshift quasars detected in PS1. Among the known quasars that do not pass our color criteria are
SDSSJ1148+5251 ($z=6.42$) and ULASJ1148+0702 ($z=6.29$) which are not strictly speaking part of our $z\sim 6$ search, 
they have $\zps - \yps > 0.5 $ and  lower limits in the $\ips-\zps$ color. 
SDSSJ1048+4637 ($z=6.20$) and ULASJ1319+0950 ($z=6.133$)
have very quasar-like colors ($\ips - \zps =2.79$ and $\zps - \yps =0.70$; and $\ips - \zps =2.42$ and $\zps - \yps =0.61$ respectively) and would be selected as 
quasars in our next search at higher redshifts (where we will allow $\zps- \yps > 0.5$). CFHQSJ1509--1749 ($z=6.12$) has only a lower limit of $\ips - \zps > 1.81$
and $z-y=0.64$, it could be detected as a quasar in our $z>6.2$ search if the $\ips$-band goes deeper in the next release of the stacked catalog.
SDSSJ2310+1855 ($z=6.0031$) has $\ips - \zps =1.91$ and $\zps - \yps =0.56$ colors and will not be selected in our next searches
because its colors are in a region with high contamination of brown dwarfs. SDSSJ0353+0104 with colors $\ips - \zps > 1.97$ and $\zps - \yps =0.46$ is 
barely missed by our selection and will be likely selected as a quasar when the depth of the $\ips$ band increases. SDSSJ2054--0005 ($z=6.0391$),
SDSSJ0927+2001 ($z=5.77$), and SDSSJ1621+5155 ($z=5.71$) are missed by our selection because their PS1 colors are hard to distinguish from the more
abundant brown dwarfs. Using our current criteria for $z \sim  6$ quasars we estimate that we recover $\sim 81\%$ (17/21) of the quasars that we 
are supposed to find ($z-y < 0.5$, excluding the PS1 quasars).  With this work, PS1  has discovered more 
than 10\% of the $z > 5.7$ quasars published thus far. Figure \ref{fig:qsos_hist} shows the redshift distribution
of all the $z\geq 5.7$ quasars published to date, highlighting the discovery surveys.

\section{How common are weak emission line quasars?}
\label{sec:wlq}
Weak-line quasars are rare objects characterized by a flat continuum and the
lack of strong emission lines. \cite{meu12} selected a sample of $\sim 1000$ quasars
($0.6 < z < 4.3$) with unusual spectra from the SDSS DR7. From these peculiar objects, they found that $18\%$ were weak-line quasars.
\cite{dia09} studied the SDSS DR5 quasar catalog and defined weak-line quasars as the ones that have the rest-frame equivalent-width (EW)
of the Ly$\alpha+$\nv\ line (determined between $\lambda_{\ensuremath{\,\rm{rest}}} = 1160\,$\AA\ and   $\lambda_{\ensuremath{\,\rm{rest}}} = 1290\,$\AA)
lower than $15.4\,$\AA. They showed that the fraction of weak-line quasars increased from $1.3\%$ at
$z< 4.2$ to $6.2\%$ at $z>4.2$. 
From Figure \ref{fig:spectra}, we noticed that half of the quasars discovered in this paper
(\Qb, \Qf, \Qg, \Qh) show a weak or
absorbed \lya\ line. 
However, following the \cite{dia09} hard-cut definition, only \Qf\ and \Qg, belong to their weak-line quasar classification with Ly$\alpha+$\nv\ EWs of 11.8 
and 10.7 \AA\ respectively. We noted that the EW is very dependent on the continuum fit estimate and since most of our spectra do not 
cover the region with $\lambda_{\ensuremath{\,\rm{rest}}}> 1500\,$\AA, a good fit to the 
continuum is challenging. Thus, the uncertainties
in our EW estimates are of the order of 25\%. Future NIR spectra of these quasars will improve these measurements.

 Even though these are low number statistics, 25\% of the quasars presented in this paper are weak-line quasars, which
is a higher fraction than found in lower redshift studies and is consistent with the 
$\sim$20\% of weak-line quasars found in
the SDSS main $z\sim 6$ quasar sample (X. Fan et al., in preparation.).

In principle, this kind of object should be easier to select at higher redshifts due to
the strong \lya\ forest and Lyman limit systems that produce a characteristic break in 
the colors of high-redshift quasars, independent of their emission lines. 
Thus, while high redshift searches based on colors---like this work---should be equally sensitive to both weak-line emission quasars and
normal quasars, some weak-line quasars
could have been missed at lower redshift due to the color-based
selection criteria.

Several scenarios have been proposed to explain the existence of such weak-line objects, including that they could be strongly lensed galaxies, BL Lac objects,
objects where the quasar activity has just started, or invoking unusual broad-line region properties in comparison to normal quasars
\citep[e.g.,][]{hry10, she10, lane11}. Nevertheless, no consensus has yet been reached.
Recently, \cite{lao11} proposed that weak-line or lineless quasars may be produced by cold accretion discs that imply non-ionizing continuum for 
some combinations of black hole masses and quasar luminosities. They claimed that very high masses in luminous active galactic nuclei are required in order
to have a cold accretion disc ($M \gtrsim 3 \times 10^9 \msun$, $L \approx 10^{46}\, \mbox{erg s}^{-1}$, especially for non-rotating black holes).
These numbers are similar to the masses and luminosities found in $z\sim 6$ quasars \citep[e.g.,][]{kur07, der11}, so this mechanism could be a simple explanation
of the high fraction of weak-line quasars in our sample.
A more extensive sample of quasars is urgently needed to shed light on the nature of these objects.

\section{Summary}
\label{sec:summary}
We have presented the discovery of eight new quasars at $z\sim 6$. With this work,
PS1 has now discovered a total of nine quasars at $5.7 \leq z \leq 6.0$.
We estimated that using the selection strategy of this paper,  we recover $\sim 81\%$ of the 
known quasars in our target redshift range (excluding the PS1 quasars).
The others are missed because their PS1 colors are hard to distinguish from brown dwarfs.
Follow-up observations are still on-going and the discovery of more quasars is expected, therefore
conclusions on the luminosity function and the space density of $z\sim 6$ quasars are not possible at this time.
The variety of spectral features among these quasars is remarkable, including four quasars with 
very bright emission lines and another four quasars
with almost no detectable emission lines.  The fraction of weak-line emission quasars found
in this work (25\%) is much larger than fractions found by other studies at lower redshifts \citep[e.g.,][]{dia09} but consistent
with the fraction in the SDSS main $z\sim 6$ quasar sample (X. Fan et al., in preparation). 
Our new discoveries show that 
weak-line emission quasars could be more common at the highest redshifts than previously thought.

\acknowledgments
E.B. thanks B. Goldmann for providing the PS1 colors of brown dwarfs, E. Schlafly for helpful discussions about PS1 data,
N. Crighton and K. Rubin for their help with MODS data reduction, C.-H. Lee and L. Johnsen for useful comments on the paper, and the IMPRS for Astronomy \& Cosmic Physics at the University of Heidelberg.
XF and IM acknowledge support from NSF grant AST 08-06861 and AST 11-07682.

The Pan-STARRS1 Surveys (PS1) have been made possible through contributions of the Institute for Astronomy, the University of Hawaii, the Pan-STARRS Project Office, the Max-Planck Society and its participating institutes, the Max Planck Institute for Astronomy, Heidelberg and the Max Planck Institute for Extraterrestrial Physics, Garching, The Johns Hopkins University, Durham University, the University of Edinburgh, Queen's University Belfast, the Harvard-Smithsonian Center for Astrophysics, the Las Cumbres Observatory Global Telescope Network Incorporated, the National Central University of Taiwan, the Space Telescope Science Institute, the National Aeronautics and Space Administration under grant No. NNX08AR22G issued through the Planetary Science Division of the NASA Science Mission Directorate, the National Science Foundation under grant No. AST-1238877, the University of Maryland, and Eotvos Lorand University (ELTE).

This work is based on observations made with ESO Telescopes at the La Silla Paranal Observatory under programmes ID 088.A-0119, 088.A-9004,
089.A-0290, 090.A-0383, 090.A-0642, and 091.A-0421.

The LBT is an international collaboration among institutions in the United States, Italy and Germany. The LBT Corporation partners are: The University of Arizona on behalf of the Arizona university system; Istituto Nazionale di Astrofisica, Italy;  LBT Beteiligungsgesellschaft, Germany, representing the Max Planck Society, the Astrophysical Institute Potsdam, and Heidelberg University; The Ohio State University; The Research Corporation, on behalf of The University of Notre Dame, University of Minnesota and University of Virginia.

This paper used data obtained with the MODS spectrographs built with
funding from NSF grant AST-9987045 and the NSF Telescope System
Instrumentation Program (TSIP), with additional funds from the Ohio
Board of Regents and the Ohio State University Office of Research.

Part of the funding for GROND (both hardware as well as personnel) was 
  generously granted from the Leibniz-Prize to Prof. G. Hasinger 
  (DFG grant HA 1850/28-1).

We acknowledge the use of the Calar Alto Faint Object Spectrograph at the 2.2m telescope (CAFOS) and MMT/SWIRC, for 
follow-up of some of the early PS1 candidates.

Based on observations collected at the Centro Astron\'omico Hispano Alem\'an (CAHA) at Calar Alto,
operated jointly by the Max-Planck Institut f\"ur Astronomie and the Instituto de Astrof\'isica de Andaluc\'ia (CSIC)'

Observations reported here were obtained at the MMT Observatory, a joint facility of the University of Arizona and the Smithsonian Institution.

This paper includes data gathered with the 6.5 meter Magellan Telescopes located at Las Campanas Observatory, Chile.

This publication makes use of data products from the Two Micron All Sky Survey,
which is a joint project of the University of Massachusetts and the Infrared 
Processing and Analysis Center/California Institute of Technology, funded by 
the National Aeronautics and Space Administration and the National Science Foundation.

Funding for SDSS-III has been provided by the Alfred P. Sloan Foundation, the Participating Institutions, the National Science Foundation, and the U.S. Department of Energy Office of Science. The SDSS-III web site is http://www.sdss3.org/.

SDSS-III is managed by the Astrophysical Research Consortium for the Participating Institutions of the SDSS-III Collaboration including the University of Arizona, the Brazilian Participation Group, Brookhaven National Laboratory, University of Cambridge, Carnegie Mellon University, University of Florida, the French Participation Group, the German Participation Group, Harvard University, the Instituto de Astrofisica de Canarias, the Michigan State/Notre Dame/JINA Participation Group, Johns Hopkins University, Lawrence Berkeley National Laboratory, Max Planck Institute for Astrophysics, Max Planck Institute for Extraterrestrial Physics, New Mexico State University, New York University, Ohio State University, Pennsylvania State University, University of Portsmouth, Princeton University, the Spanish Participation Group, University of Tokyo, University of Utah, Vanderbilt University, University of Virginia, University of Washington, and Yale University.

This publication makes use of data products from the \textit{Wide-field Infrared Survey Explorer},
which is a joint project of the University of California, Los Angeles, and the Jet Propulsion Laboratory/California Institute of Technology,
funded by the National Aeronautics and Space Administration.

This research made use of Astropy, a community-developed core Python package for Astronomy \citep[][\url{http://www.astropy.org}]{astropy13}.
This publication made use of TOPCAT  \citep[][\url{http://www.starlink.ac.uk/topcat}]{tay05} and 
STILTS \citep[][\url{http://www.starlink.ac.uk/stilts}]{tay06}.
The plots in this publication 
were produced using Matplotlib \citep[][\url{http://www.matplotlib.org}]{hun07}.

{\it Facilities:} \facility{PS1 (GPC1)},\facility{VLT:Antu (FORS2)}, \facility{NTT (EFOSC2)}, \facility{LBT (MODS)}, \facility{Magellan:Baade (FIRE)},
\facility{CAO:3.5m (Omega2000)}, \facility{CAO:2.2m (CAFOS)}, \facility{Max Planck:2.2m (GROND),  \facility{MMT (SWIRC)}}

\appendix

\section{Pan-STARRS1 low-quality flags} \label{ap:ps1_flags}

Here we present Table \ref{table:flags}, containing the PS1 flags used in our selection.
\begin{table*}[htdp]
\caption{Pan-STARRS1 Bit-flags Used to Exclude Bad or Low-quality Detections. }
\begin{center}
\begin{tabular}{lll}
\hline
\hline
{\bf FLAG1 NAME} & {\bf Hex Value} & {\bf Description} \\
\hline
FITFAIL  &  0x00000008 & Fit (nonlinear) failed (non-converge, off-edge, run to zero) \\
POORFIT 	   &  0x00000010 & Fit succeeds, but low-SN or high-Chisq \\
PAIR  &  0x00000020 & Source fitted with a double psf \\
SATSTAR	   &  0x00000080 & Source model peak is above saturation \\
BLEND 	   & 0x00000100  & Source is a blend with other sources \\
BADPSF  & 0x00000400 & Failed to get good estimate of object's PSF \\
DEFECT  & 0x00000800 & Source is thought to be a defect \\
SATURATED  & 0x00001000 & Source is thought to be saturated pixels (bleed trail) \\
CR\_LIMIT  & 0x00002000 & Source has crNsigma above limit \\
EXT\_LIMIT  & 0x00004000 & Source has extNsigma above limit \\
MOMENTS\_FAILURE  & 0x00008000 & Could not measure the moments \\
SKY\_FAILURE  & 0x00010000 & Could not measure the local sky \\
SKYVAR\_FAILURE  & 0x00020000 & Could not measure the local sky variance\\
MOMENTS\_SN  & 0x00040000 & Moments not measured due to low S/N\\
BLEND\_FIT  & 0x00400000 & Source was fitted as a blend\\
SIZE\_SKIPPED  & 0x10000000 & Size could not be determined\\
ON\_SPIKE  & 0x20000000 & Peak lands on diffraction spike\\
ON\_GHOST  & 0x40000000 & Peak lands on ghost or glint\\
OFF\_CHIP  & 0x80000000 & Peak lands off edge of chip\\
\hline
\hline
{\bf FLAG2 NAME} & {\bf Hex Value} & {\bf Description} \\
\hline
ON\_SPIKE	& 0x00000008	& $> 25\%$ of pixels land on diffraction spike	\\
ON\_STARCORE	& 0x00000010	& $> 25\%$ of pixels land on star core	\\
ON\_BURNTOOL	& 0x00000020	& $> 25\% $of pixels land on burntool subtraction region\\
\hline
\end{tabular}
\end{center}
\label{table:flags}
\end{table*}

\section{ \Qa\ Submillimeter observations} \label{ap:submm}
 
\begin{figure}
\epsscale{0.8}
\plotone{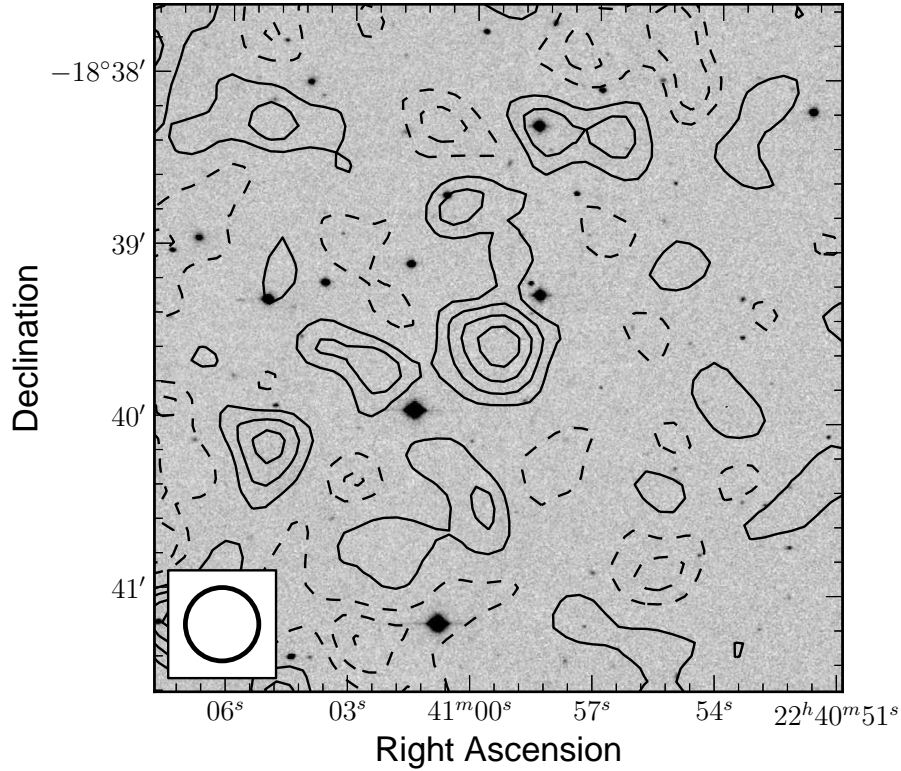}
\caption{Image centered on the serendipitous 870~$\mu$m source at a distance of $2.\arcmin5$ from \Qa\ (outside the image).
The solid (dashed) lines are the positive (negative) $1\sigma$-contours overlaid over the \Jgrond\ image.
The detection is $9.50 \pm 2.02$ mJy. There is no evident optical counterpart.
The LABOCA beam size of $18\sec 2 \times 18\sec 2 $ is shown in the bottom left panel.\label{fig:submm}}
\vspace{2.mm}
\end{figure}

 The far-infrared (FIR) emission traces the warm dust emission from quasar host galaxies and allows us to estimate their star formation rates.
Studies from $z\sim 6$ quasars have shown that $\sim 30\%$ of these quasars are bright at millimeter and submillimeter wavelengths \citep[e.g.,][]{wan08}.
With this in mind, we took 870~$\mu$m observations of \Qa\ using the Large APEX Bolometer Camera \citep[LABOCA;][]{sir09} on the 12 m APEX telescope
\citep{gus06}. The observations were carried out during 2012 November for a total of 14.5 hr. The quasar was observed mainly in the afternoon 
or early evening in mediocre to poor weather conditions. The data were reduced using the standard 
procedures implemented in the BoA software \citep{schu12}.
The quasar remained undetected in the map with an rms of 1.6 mJy 
at the quasar position.
In the same map, however, we detected a tentative $4.7\sigma$ submillimeter source. Its coordinates are R.A.= 22:40:59.407 and Decl.=-18:39:34.94.
It is located at a distance of $\sim 2.\arcmin5$ from \Qa\ and we did not find an evident optical counterpart. 
The 870~$\mu$m flux of this serendipitous source is $9.50 \pm 2.02$ mJy. Figure \ref{fig:submm} shows the 870~$\mu$m  $1\sigma$-contours overlaid
over the \Jgrond\ image.

\section{\textit{WISE} magnitudes} \label{ap:wise}

We cross-matched the PS1 quasars from Table \ref{table:ps1qso_stackprop} with the \textit{WISE} All-Sky data release products catalog (Cutri et al. 2012) within a radius
of 3\arcsec. Three quasars were detected in the main all-sky release source catalog with S/N $> 5.0$ in the W1 band and two quasars were detected in the 
\textit{WISE} `Reject Table' \footnote{\url{http://wise2.ipac.caltech.edu/docs/release/allsky/expsup/sec2\_4a.html}} with S/N $> 3.0$ in the W1 band. The \textit{WISE}
photometry is presented in Table \ref{table:wise}.
For completeness, in Table \ref{table:wise} we also included the \textit{WISE} photometry for the \textit{WISE}-detected quasars from Tables \ref{table:qsos_other}
and \ref{table:qso_no_sel}.

\begin{table*}[htdp]
\caption{\textit{WISE} Magnitudes for \textit{WISE}-detected Quasars from Table \ref{table:ps1qso_stackprop} (the five entries at the top), Table \ref{table:qsos_other}
(the 17 entries in the middle), and Table \ref{table:qso_no_sel} (the nine entries at the bottom).}
\begin{center}
\begin{tabular}{lccccccccc}
\hline
\hline
QSO 		      &    W1    & $\sigma_1$  &  W2     & $\sigma_2$& W3       & $\sigma_3$ & W4      & $\sigma_4$ & Table\tablenotemark{a}\\
\hline
PSO J340.2041--18.6621 &	 $16.744$& $0.143$	& $15.851$& $0.231$  & $12.555$ & $\ldots$   &  $9.060$& $\ldots$   & M   \\
PSO J007.0273+04.9571 &	 $17.606$& $0.286$	& $16.194$& $0.310$  & $12.403$ & $\ldots$   &  $8.628$& $\ldots$   & R \\
PSO J037.9706--28.8389 & $18.025$& $0.339$	& $16.454$& $\ldots$  & $12.913$ & $\ldots$   &  $8.995$& $\ldots$   & R \\
PSO J183.2991--12.7676 & $16.387$& $0.098$	& $16.323$& $0.310$  & $12.710$ & $\ldots$   &  $8.625$& $\ldots$   & M \\
PSO J215.1514--16.0417 &	 $15.567$& $0.047$	& $14.778$& $0.068$  & $11.782$ & $0.205$    &  $8.758$& $0.329$   & M \\
\hline
SDSSJ1030+0524\tablenotemark{b}  &	$16.512$& $0.114$	& $15.567$& $0.174$  & $12.364$  & $\ldots$  &  $8.372$& $\ldots$   & M \\
SDSSJ1623+3112\tablenotemark{b}  & 	$16.839$& $0.110$	& $15.914$& $0.166$  & $12.706$  &  $\ldots$  & $9.268$& $\ldots$   & M\\
SDSSJ1250+3130\tablenotemark{b}  & 	$16.489$& $0.096$	& $15.474$& $0.136$  & $12.302$  &  $\ldots$  & $8.473$& $\ldots$   & M\\
SDSSJ1602+4228\tablenotemark{b}  & 	$16.107$& $0.046$ 	& $15.209$ & $0.062$ & $12.184$  & $0.184$	& $9.526$& $\ldots$  & M\\
SDSSJ1630+4012\tablenotemark{b}  & 	$18.000$& $0.271$	& $17.138$ & $\ldots$& $13.122$  & $\ldots$  &  $9.447$& $\ldots$    & R  \\
ULASJ1207+0630  & 	$17.217$& $0.209$	& $16.005$ & $0.265$ & $12.514$  & $\ldots$  &  $8.442$& $\ldots$    & M\\
SDSSJ1137+3549\tablenotemark{b}  &	$16.379$& $0.092$	& $15.868$ & $0.193$  & $12.144$ & $\ldots$  &  $8.765$& $\ldots$    & M\\
ULASJ0148+0600  & 	$16.200$& $0.066$	& $15.267$ & $0.099$  & $12.753$ & $\ldots$  &  $8.728$& $\ldots$    & M \\
SDSSJ1335+3533  & 	$16.936$& $0.126$	& $15.944$ & $0.170$  & $12.756$ & $\ldots$  &  $9.405$& $\ldots$    & M  \\
SDSSJ1411+1217  & 	$16.709$& $0.088$	& $15.612$ & $0.107$  & $13.014$ & $\ldots$  &  $9.375$& $\ldots$    & M  \\
SDSSJ0005--0006  & 	$17.707$& $0.319$	& $16.827$ & $\ldots$  & $12.368$ & $\ldots$  &  $8.359$& $\ldots$    & R  \\
NDWFSJ1425+3254 & 	$17.147$& $0.144$	& $16.607$ & $0.241$  & $13.288$ & $\ldots$  &  $9.083$& $\ldots$    & M\\
ULASJ1243+2529  & 	$16.72$	& $0.144$	& $15.594$ & $0.149$  & $12.826$ & $\ldots$  &  $9.229$& $\ldots$    & M \\
SDSSJ1436+5007  & 	$17.541$& $0.165$	& $16.771$ & $0.271$  & $13.203$ &$\ldots$  &  $9.532$& $\ldots$    & M \\
SDSSJ0002+2550  & 	$16.330$& $0.078$	& $15.410$ & $0.147$  & $12.012$ & $0.259$	 &  $8.579$& $\ldots$    & M \\
SDSSJ1044--0125  & 	$16.274$& $0.088$	& $15.546$ & $0.155$  & $12.281$  &$0.362$  &  $9.532$& $\ldots$    & M  \\
ULASJ0203+0012  &	$16.656$& $0.095$     & $16.546$ & $0.293$  & $12.381$  &$0.339$  &  $9.369$& $\ldots$    & M \\
\hline
SDSSJ1148+5251\tablenotemark{b}  &	$16.007$& $0.062$     & $15.242$ & $0.093$  & $12.544$  &$0.350$  &  $8.598$& $\ldots$    & M \\
ULASJ1148+0702  &	$16.632$& $0.131$     & $15.537$ & $0.167$  & $12.534$  &$\ldots$  &  $8.700$& $\ldots$    & M \\
SDSSJ1048+4637\tablenotemark{b}  & 	$16.430$& $0.080$     & $16.259$ & $0.233$  & $12.885$  &$\ldots$  &  $8.843$& $\ldots$    & M \\
ULASJ1319+0950\tablenotemark{b}  & 	$17.222$& $0.145$     & $16.848$ & $0.400$  & $12.966$  &$\ldots$  &  $9.030$& $\ldots$    & M \\
SDSSJ0353+0104\tablenotemark{b}  & 	$16.886$& $0.147$   & $16.497$ & $0.375$  & $12.225$  &$\ldots$  &  $8.560$& $\ldots$    & M \\
SDSSJ2054--0005\tablenotemark{b}	  & 	$18.017$& $0.339$   & $16.250$ & $\ldots$  & $12.595$  &$\ldots$  &  $8.727$& $\ldots$    & R \\
SDSSJ2310+1855  &	$15.950$& $0.067$    & $15.192$ & $0.107$  & $12.541$  &$\ldots$  &  $9.003$& $\ldots$    & M \\
SDSSJ0927+2001  &	$17.134$& $0.205$    & $16.776$ & $\ldots$  & $11.994$  &$\ldots$  &  $8.423$& $\ldots$    & M \\ 
SDSSJ1621+5155  & 	$15.711$& $0.036$    & $14.782$ & $0.043$  & $13.033$  &$0.293$  &  $9.609$& $\ldots$    & M \\ 

\hline
\end{tabular}
\tablenotetext{0}{A detection is considered when a S/N greater than 3.0 in the W1 band is reported.
 Error values are only listed when S/N$ > 3.0$; otherwise null results ($\ldots$) are listed.}
\tablenotetext{1}{The \textit{WISE} table where the quasar information are found. M: Main \textit{WISE} all-sky release source catalog. R: \textit{WISE} Reject Table.}
\tablenotetext{2}{The \textit{WISE} magnitudes for these quasars were also reported by \cite{bla13}.}

\end{center}
\label{table:wise}
\end{table*}


\end{document}